\documentclass[aps,prl,showpacs,twocolumn,superscriptaddress]{revtex4}
\usepackage{epsfig}
\begin{document}


\title{Analysis of (p,$\gamma$) capture cross-sections relevant to p-process using TALYS for A=75-110}

\author{Satabdi Mondal}
\affiliation{Department of Physics, Bankura University, Bankura, West Bengal-722155, India}

\author{Enakshi Senapati}
\affiliation{DDepartment of Physics, Bankura University, Bankura, West Bengal-722155, India}

\author{Deepak Pandit}
\affiliation{Variable Energy Cyclotron Centre, 1/AF-Bidhannagar,
Kolkata, Pin-700064, WB, India}

\affiliation{Homi Bhabha National Institute, Training School Complex,
Anushaktinagar, Mumbai, Pin-400094}

\author{Balaram Dey}
\affiliation{Department of Physics, Bankura University, Bankura, West Bengal-722155, India}

\author{Pampa Das}
\affiliation{Department of Physics, West Bengal State University, Kolkata-700126, W.B., India}

\author{Alokkumar De}
\affiliation{ Ex-Department of Physics, Raniganj Girls’ College, Raniganj-713358, W.B., India.}

\author{Srijit Bhattacharya}
\affiliation{Department of Physics, Haldia Govt. College, Purba Medinipur, Pin-721657, WB, India}
\email[e-mail:]{srijit.bha@gmail.com}



\date{\today}

\begin{abstract}
The proton capture (p, $\gamma$) cross-sections for eight different atomic nuclei in the mass region A=75-110 were calculated within the nuclear reaction model code TALYS. For all the reactions, we tested different combinations of inputs for level density (l.d) parameter and  gamma strength function ($\gamma_{sf}$). {Finally, it was observed that application of hybrid input in TALYS (macroscopic l.d and microscopic or semi-microscopic $\gamma_{sf}$ or in abbreviation mac-mic) resulted successful agreement of theoretical prediction with the existing experimental data.} Isospin correction was also incorporated in a few cases which improved the matching if the centre of mass energy reaches the threshold energy of the opening of (p,n) channel. The corresponding thermonuclear reaction rates were calculated for all the nuclei and some discrepancies were found with the prediction of the NON-SMOKER code. Using the particular mac-mic input combination, the cross-section and reaction rate for the nuclei $^{92}$Nb and $^{92}$Mo are calculated within TALYS. These two nuclei lack experimental data but are highly important for understanding early solar system processes. This is a rare attempt to explain the p-capture cross-section of different p-nuclei (A=75-110 range) with similar set of input combinations in TALYS, which may help to remove the uncertainty generated due to the variation of input parameters within nuclear statistical model codes.

\end{abstract}
\maketitle

\section{Introduction}

The theory of synthesis of elements, heavier than iron, is one of the most interesting as well as puzzling topics in the field of nuclear astrophysics. The finding of completely accurate and reliable model of nucleosynthesis still remains obscure to us. Neutron capture (rapid r and slow s) reactions account for the synthesis of 99\% of the elements heavier than iron \cite{Burb,Lang}. But there are about 30-35 proton rich nuclides between $^{74}$Se and $^{194}$Hg, the origin of which could not be predicted by this neutron capture process. These nuclei are known as ”p-nuclei”. These p-nuclei are much less abundant (10-100 times) than the nuclei produced via r and s processes, albeit abundance trend lines are similar in nature. Therefore, it is thought that these p-nuclei are produced from the r or s-process seed nuclei, either by photo-disintegration reactions (also called $\gamma$-process), namely ($\gamma$,p), ($\gamma$,n) and ($\gamma$,$\alpha$), or by proton capture \cite{Boyd,Gyur,Woos,Woo1}.
To understand the p-process and the corresponding reaction rates, the knowledge of cross-section of such reactions or their inverse (related by detailed balance) is indeed necessary. The p-process involves as many as 2000 nuclei forming the reaction network of more than 10000 reactions \cite{Raye} out of which the experimental determination of cross-sections of several reactions is a difficult task due to the hurdles involved in the measurement of very low cross-sections. Therefore, experimentally, only a handful of data is available so far. 

Theoretically, estimation of cross-section depends largely on the Hauser-Feshbach (HF) model \cite{Raus,Haus} dependent codes like TALYS \cite{Koni}, EMPIRE \cite{Herm}, SMARAGD \cite{Rau1} (the modified version of NON-SMOKER \cite{Rau2}), etc. But the correct predictions by such codes depend largely on the input parameters. In absence of experimental data, the uncertainty of the input parameters can lead to a large variation in the predicted cross-section.
The highly popular HF model codes such as NON-SMOKER, TALYS and EMPIRE differ mostly in input models such as level density parameter (l.d), optical model potential (OMP) and E1 gamma strength function ($\gamma_{sf}$). In the past, the predictions of these codes showed varying results when compared with experimental (p,$\gamma$) cross-sections. In some cases, the NON-SMOKER predicted cross-sections agreed with existing experimental data better than that of TALYS and EMPIRE. On the contrary, There are several cases, TALYS and EMPIRE were found successful \cite{Spyr,Khal,Wu1} rather than NON-SMOKER. Surprisingly, there are cases, such as Khaliel et al \cite{Khal}, no unique combination of input models could be found reproducing all the experimental data points for even a single nucleus. As a result, it becomes difficult to use a code confidently to predict the reaction cross-section for a nucleus, especially for the reactions having no experimental data at all. Besides, in majority of the cases, it is not mentioned in the concerned references which particular $\gamma$ strength function model has been used. Under these circumstances, our endeavour is to explain the existing experimental data (in the mass region A=75-110) on (p,$\gamma$) cross-section in a uniform way without changing the input parameter models viz. level density parameter (l.d), optical model (OMP) and E1 gamma strength function ($\gamma_{sf}$) in TALYS. The unique combination of the input models could potentially resolve the uncertainty involved in the theoretical cross-section calculations. Further, we estimate thermonuclear reaction rates in these nuclei and also compare with existing theoretical data. 

Here we mainly deal with some p-nuclei that generally show lower abundances than the predictions of p-process nucleosynthesis models. As an example, the nuclei in the mass range A=90-96 are generally known for lower abundance than the expected value as per theoretical models and hence highly important for the determination of most favored photo-disintegration sites for p-process. It is known that the explosively burning O/Ne layers in type-II or core-collapse supernovae (SN II), producing temperatures in the range of 1.5-3.5 GK, are the most favored sites in the present-day model of the origin of p-nuclei \cite{Mei1}. But interestingly, it could not account for the predicted underproduction (by as much as a factor of 20-50) in the abundances of the nuclei $^{92,94}$Mo and $^{96,98}$Ru. This problem could be resolved either by changing the seed nuclei abundance or by assuming alternate sites (rp or $\nu$p processes) to account for the origin of p-nuclei involving (p,$\gamma$) reactions \cite{Arno,Scha,Froh}. For the latter case, the need for high proton density and temperature suggests type I supernova (SNe-Ia), x-ray bursters or the hydrogen rich outer layer of type-II supernova (SN-II) could be possible sources. But feasibility of these sites is highly debated and inconclusive \cite{Bork}. 

In this connection the importance of cross-section, reaction rate and the abundance of the nuclei $^{92}$Nb and $^{92}$Mo is enormous. $^{92}$Nb is produced by $\gamma$- process ($^{91}$Zr(p,$\gamma$)$^{92}$Nb, $^{93}$Nb($\gamma$,n)$^{92}$Nb reactions) and cannot be produced by $\nu$p or rp processes, hence it can be an ideal source to investigate the validity of different theoretical nucleosynthesis models. Moreover, until now the observed abundance of $^{92}$Nb cannot be predicted correctly indicating the shortcomings of existing theoretical models of nucleosynthesis. The abundance ratio of $^{92}$Nb/$^{92}$Mo is usually estimated using these models owing to the reason that $^{92}$Mo is a nucleus originated by p-process only. This is discussed in detail in ref.\cite{Gyur}. $^{92}$Nb has ground state with very long life time (half life of 3.47x10$^7$ years) and hence only the cross-section of its isomeric state can be measured in experiment for verification. This study has been done very recently \cite{Lizu}. Again, rarely has an attempt been made to measure the cross-section for the p-capture reaction producing $^{92}$Mo. Only the $\gamma_{sf}$ and thermonuclear rate were measured in a recent work \cite{Tvet}. Therefore, we calculate the cross-section and reaction rate of the reactions $^{91}$Zr(p,$\gamma$)$^{92}$Nb and $^{91}$Nb(p,$\gamma$)$^{92}$Mo with TALYS. For this calculation, we select the input models in TALYS by rigorously testing each model on eight different nuclides within A=75-110 region having existing experimental data.

\section{Data extraction}\label{sec2}
The experimental cross-sections for (p,$\gamma$) reaction from the daughter nuclides $^{75}$As,  $^{84}$Sr, $^{91,93}$Nb, $^{93,99}$Tc and  $^{108,110}$Cd were extracted from references: \cite{Wu1}, \cite{Lota} and \cite{Will},  \cite{Spyr},  \cite{Gyur}, \cite{Khal}, respectively.
{In-beam measurement of $\gamma$-rays was performed in the reaction $^{74}$Ge(p,$\gamma)^{75}$As by Wu et al. \cite{Wu1} within E$_{cm}$= 1.2-3.7 MeV. They measured the $\gamma$-angular distribution using 4 closely mounted HPGe detectors.  A similar method was adopted in the $\gamma$-ray measurements for the reactions $^{107,109}$Ag(p,$\gamma$)$^{108,110}$Cd \cite{Khal}. The measurement of $\gamma$-rays coming from thick targets irradiated with proton beams was performed in the experiments $^{92,98}$Mo(p,$\gamma$)$^{93,99^m}$Tc  \cite{Gyur}. Another method was followed in the reactions $^{90,92}$Zr(p,$\gamma$)$^{91,93}$Nb studied by Spyrou et al \cite{Spyr}. They used a NaI(Tl) total absorption spectrometer to measure the average multiplicity of the $\gamma$-rays coming from the reactions in the proton energy range 2.0-5.2 MeV. They revisited the earlier experiment performed by Laird et al \cite{Lair}. In a unique experiment performed at the ISAC-II facility of TRIUMF, for the reaction $^{83}$Rb(p,$\gamma$)$^{84}$Sr, the radioactive ion beam of $^{83}$Rb was utilized to populate the $^{84}$Sr nucleus and using the $\gamma$-ray array TIGRESS, the cross section of the reaction was measured \cite{Lota}.} We chose all these experimental datasets because these data points lie well within the astrophysically significant Gamow energy window region. The energy window mentioned represents the thermonuclear temperature range of 1.5-3.5 GK that is actually the typical temperature for proton capture reactions inside the stellar environment \cite{Mei1,Chen}
 
\section{Theoretical model}
{In this paper, we use the TALYS nuclear reaction code (version 2.0) to simulate the nuclear reactions}. This code uses Hauser-Feshbach (HF) model to calculate the equilibrium particle emission. This code provides the opportunity to use different microscopic as well as phenomenological optical model potentials. In addition, different options on the level density and $\gamma_{sf}$ may also be selected by the user. In this work, our search for the best possible level density and $\gamma_{sf}$ models for the nuclear mass region A=75-110 begins with the comparison of extracted experimental data from the literature with TALYS predictions. The matching between extracted experimental data and theoretical predictions is done by changing input combinations, and best fit input parameters are obtained for each nuclei, separately. In this process the l.d models are varied between TALYS in-built option 1 (CT or constant temperature Fermi gas model \cite{Gilb}), option 2 (BSFG or back shifted Fermi gas model \cite{Dilg,Koni1}, option 3 (GSM or generalized superfluid model \cite{Igna}) and microscopic models option 4-6 (microscopic level densities-4 \cite{Gor1}, 5 \cite{Gor2}, and 6 \cite{Hila}. The $\gamma_{sf}$ is varied between option 1 (BAL or Brink-Axel \cite{Brin}), option 2 (KU or Kopecky-Uhl \cite{Kope}), option 3-9 (microscopic models- 3 and 4 \cite{Gor3}, 5 \cite{Gor4}, 6-8 \cite{Koni,Mart,Gor5,Gor6} and semi-microscopic model-9\cite{Gor7}). {Option 8 is microscopic axially deformed Hartree-Fock-Bogolyubov (HFB) plus quasiparticle random-phase approximation (QRPA) based on the Gogny D1M  interaction. Option 9 is SMLO (simple modified
 E1 Lorentzian and simple M1 Lorentzian). Both these options are new addition in TALYS-2.0 code from its predecessor. In this way, we have found the best possible input model combination (l.d and $\gamma_{sf}$) that can predict the cross-section in majority of the nuclei within the mass range of our interest. Finally, such selected input models are applied on the $^{92}$Nb and $^{92}$Mo nuclei, where experimental data are scarce or do not exist at all, to find the cross-section and reaction rate.} 
In this paper a particular TALYS calculation will henceforth be called as TALYS-(x,y) where x=l.d.parameter option and y=$\gamma$ strength function option. The OMP is also varied between local and global models, switching on and off the semi-microscopic Jeukenne-Lejeune-Mahaux (JLM) model \cite{Baug}.  
The calculation is rechecked by changing the default value of the paramater 'fiso' that changes the cross-section and reaction rate. In particle capture (proton or alpha), the standard practice to incorporate isospin effects in TALYS is to reduce the $\gamma_{sf}$ by multiplying it with a small non-zero parameter fiso \cite{Lars}. TALYS does not have the advantage of using isospin-dependent l.d. model that NON-SMOKER has as a built-in input parameter. However, there is no systematic procedure how to choose fiso other than by directly comparing with the experimental data. One can also impart the same effect by reducing the input parameter 'gamgamadjust' for decreasing the $\gamma_{sf}$ without changing the parameter fiso. 'gamgamadjust' changes the $\gamma$-width and consequently changes its cross-section.
The calculation is done for the default mass table 2 of TALYS (Goriely HFB-Skyrme table).  
The reaction rates are also estimated between thermonuclear temperatures T=1-5 GK for all the nuclei using the same input models used in the determination of cross-section.

\section{RESULTS AND DISCUSSIONS}
We have collected three experimental datasets of the reaction $^{74}$Ge(p,$\gamma$)$^{75}$As \cite{Wu1,Quin,Saur}. 
Quinn et al used TALYS code with different level density models such as BSFG, CT and some microscopic models along with JLM optical model but could not reproduce the experimental data. The best matching was shown by BSFG model, although there was a disagreement of almost 40\% at the lowest energy E$_{cm}$=1.554 MeV. A recent measurement done by Wu et al. \cite{Wu1} reconfirmed the results described by Quinn et al. upto E$_{cm}$=2.76 MeV. Wu et al. showed that EMPIRE calculations with the enhanced genaralized superfluid model (EGSM) of the level density could successfully predict their results at lower energies, though some discrepancies (nearly 20\%) were observed at higher energies (3.8-4.1 MeV).  
 
We performed the TALYS calculation with all the existing options of l.d and $\gamma_{sf}$ for $^{75}$As nucleus. We also simultaneously calculated the maximum deviation and the reduced $\chi^2$ values using the equation,
\begin{equation}
\chi^2 =\sum_{i=1}^{n}{[O_i-E_i]}^2/E_i
\end{equation}
where, O$_i$ and E$_i$ are the observed (calculated using the code) value and expected (from experimental data) value, respectively. The performance of hybrid TALYS-(2,5), TALYS-(2,8) and TALYS-(2,9), using macroscopic l.d. model and microscopic (and also the semi-microscopic option SMLO) $\gamma_{sf}$ (hereafter we term it as mac-mic combination) input, is better than any other combinations to reproduce the experimental data, even better than the microscopic l.d model and microscopic $\gamma_{sf}$ (mic-mic) option. The combination TALYS-(2,5) with local OMP along with isospin correction factor 'fiso' of 0.6 gives minimum value of $\chi^2$ and deviation in the matching of the experimental data.

\begin{figure}
\begin{center}
\includegraphics[height=8.0 cm, width=8.0cm]{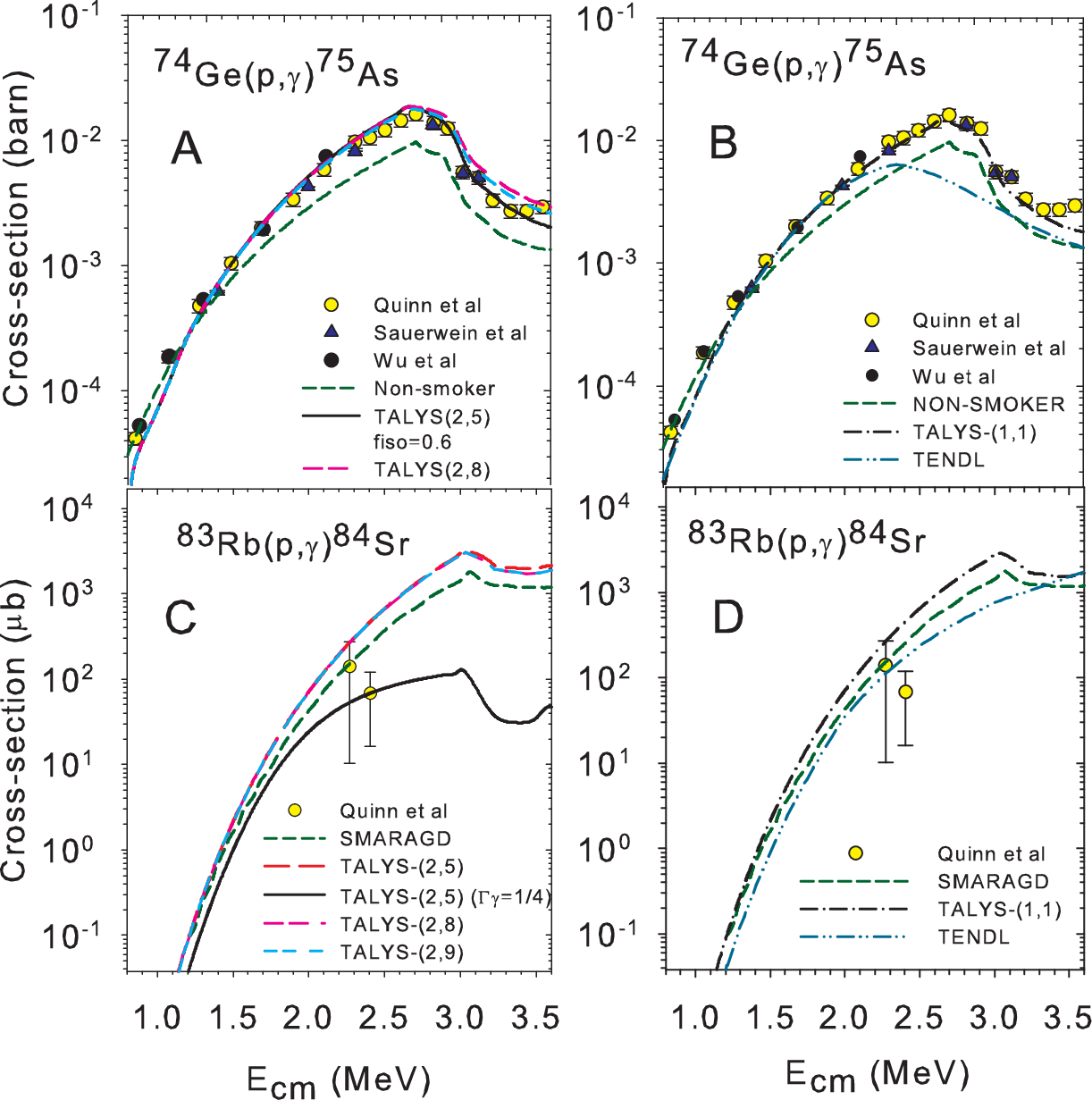}
\caption{\label{fig1} (Online color) The cross-sections (shown by markers) for the nuclei $^{75}$As and $^{84}$Sr are plotted against the centre of mass energy E$_{cm}$ of the system. Panel A:  TALYS input predictions (mac-mic) for the nucleus $^{75}$AS are shown. TALYS-(2,5) with isospin correction fiso=0.6, TALYS-(2,8) and TALYS-(2,9) using black continuous line, pink medium-medium dashed-dashed curve and cyan short-short dashed-dashed curve are depicted,respectively. The green short-short dashed curve is the NON-SMOKER prediction. Panel B: TALYS predictions with mac-mac inputs are shown along with experimental data symbols.  Green long dashed curve is the NON-SMOKER calculation and black dashed-dotted curve is the TALYS-(1,1) calculation. TENDL plot is also shown using dashed-dotted-dotted line of dark cyan color in this panel. Panels C and D: Cross-section plot of $^{84}$Sr is shown along with TALYS mac-mic inputs (panel C) and mac-mac inputs (panel D), respectively.}
\end{center}
\end{figure}

However, TALYS-(2,5) cross-section predictions could not match the data below E$_{cm}$=1.94 MeV, where in our calculation 25\% overprediction is observed. Other TALYS mic-mic inputs also give the same results at such energies. However, this agreement is much better than the earlier TALYS calculation (40\% underprediction) reported by Wu et al. Therefore, our calculation gives an overall satisfactory result starting from lowest to the highest values of center of mass energies. The top panels (A and B) of Fig.\ref{fig1} show the experimental data points of cross-section (in $^{75}$As) indicated by yellow filled circles (Quinn et al.), blue open triangles (Sauerwein et al.) and black filled squares (Wu et al.). Panel A gives TALYS with mac-mic inputs while panel B shows TALYS with mac-mac inputs.  The black continuous line, pink medium-medium and cyan short-short dashed curves represent TALYS-(2,5) outputs with isospin correction (fiso=0.6), TALYS-(2,8) and TALYS-(2,9), respectively. Panel B gives TALYS-(1,1) (black dashed-dotted line) and TENDL 23 (TALYS based evaluated nuclear data library version 23) predictions. These predictions do not agree with the experimental data. Both panels contain NON-SMOKER prediction (green short dashed line) that gives better result upto E$_{cm}$=1.95 MeV and after that shows large deviations from the experiment.

\begin{figure}[htbp]
\begin{center}
\includegraphics[height=8.0 cm, width=8.0cm]{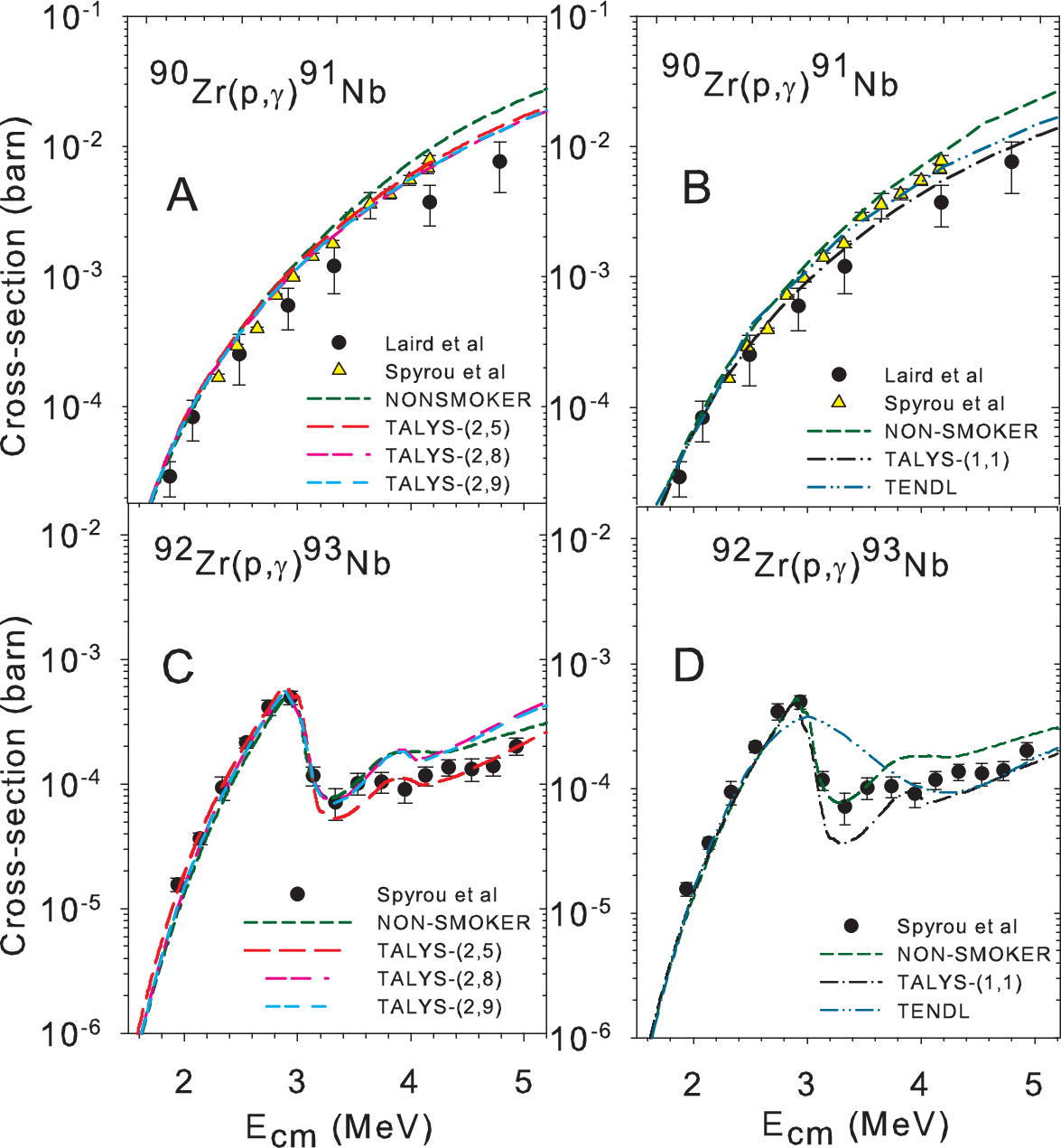}
\caption{\label{fig2} (Online color) Panels A and B: TALYS predictions with mac-mic and mac-mac inputs, respectively, for $^{91}$Nb. Yellow open circles represent data of Laird et a.l and black filled circles gives data of Spyrou et al. TALYS-(2,5), TALYS(2,8), TALYS-(2,9) are shown by red long dashed, pink medium-medium dashed and short-short dashed curve, respectively, in panel A and TALYS-(1,1) by dashed-dotted curve in panel B. Panels C and D: TALYS predictions with mac-mic and mac-mac inputs are demonstrated. In panel C, TALYS-(2,5) with fiso=0.6, TALYS-(2,8) and TALYS-(2,9) are shown by black continuous line, pink medium-medium dashed curve and cyan short-short dashed cuve, respectively. In panel D, TALYS-(1,1) is depicted by dashed-dotted curve and TENDL output is given by dark cyan dashed-dotted-dotted curve. NON-SMOKER plot is shown by green short dashed curve in both the panels.}
\end{center}
\end{figure}

\begin{figure}[htbp]
\begin{center}
\includegraphics[height=8.0 cm, width=8.0cm]{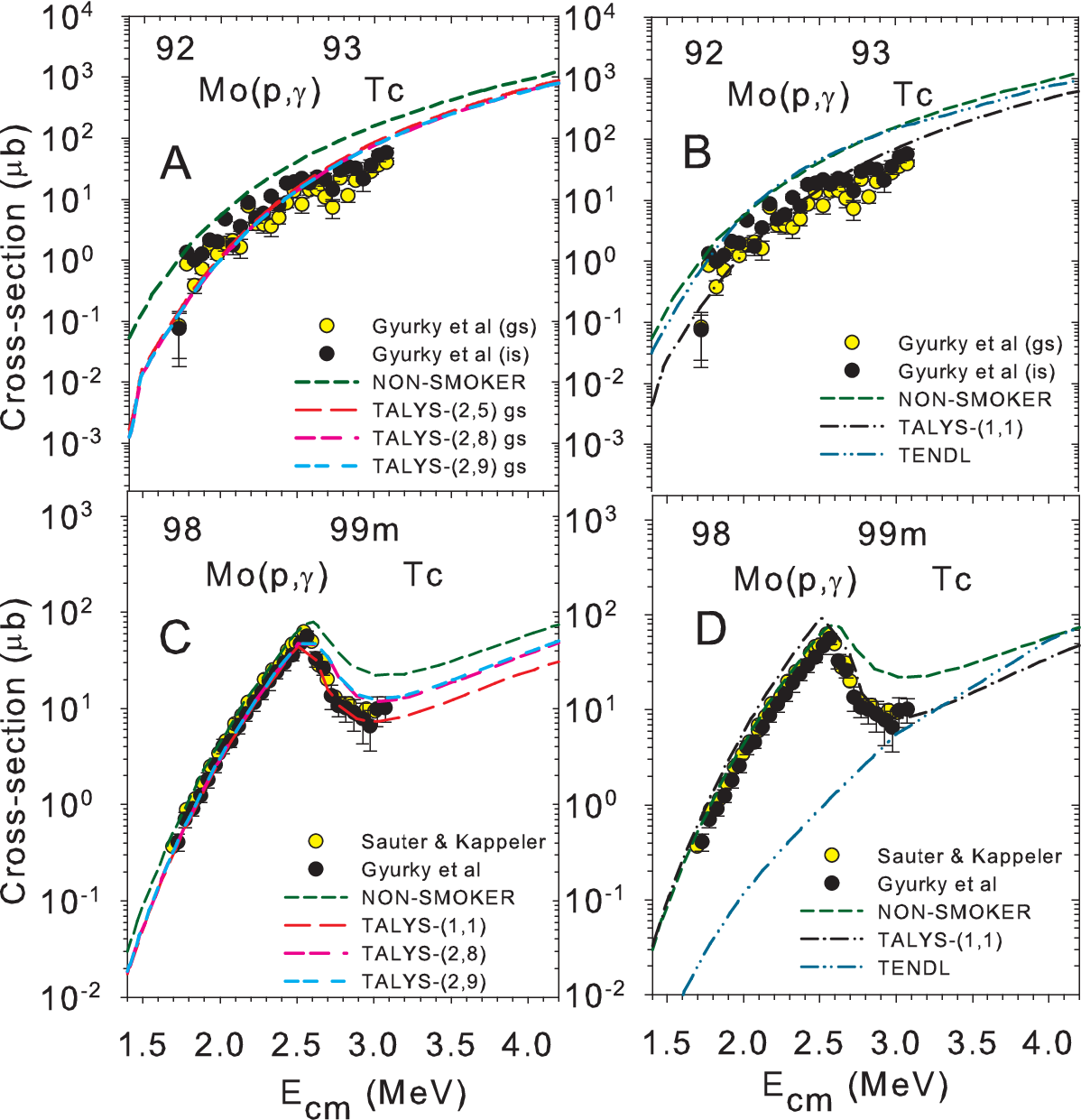}
\caption{\label{fig3} (Online color) Panels A and B: TALYS predictions with mac-mic and mac-mac inputs, respectively, for $^{93}$Nb. Yellow open circles represent experimental data of Gyurky et al for ground state of $^{93}$Tc and black filled circles are the experimental data of Gyurky et al for isomeric states of $^{93}$Tc. TALYS-(2,5), TALYS(2,8), TALYS-(2,9), TALYS-(1,1) ground state predictions (mac-mic inputs), and TENDL are also shown as per the conventions followed by the earlier figures in this manuscript. In both panels, NON-SMOKER calculations are also demonstrated. Panels C and D: TALYS predictions with mac-mic and mac-mac inputs are demonstrated for $^{99m}$T. In panel C, TALYS-(2,5) with fiso=0.6, TALYS-(2,8) and TALYS-(2,9) are shown. In panel D, TALYS-(1,1) is depicted along with TENDL output. In both the panels, NON-SMOKER plot is also shown.}
\end{center}
\end{figure}

\begin{figure}[htbp]
\begin{center}
\includegraphics[height=8.0 cm, width=8.0cm]{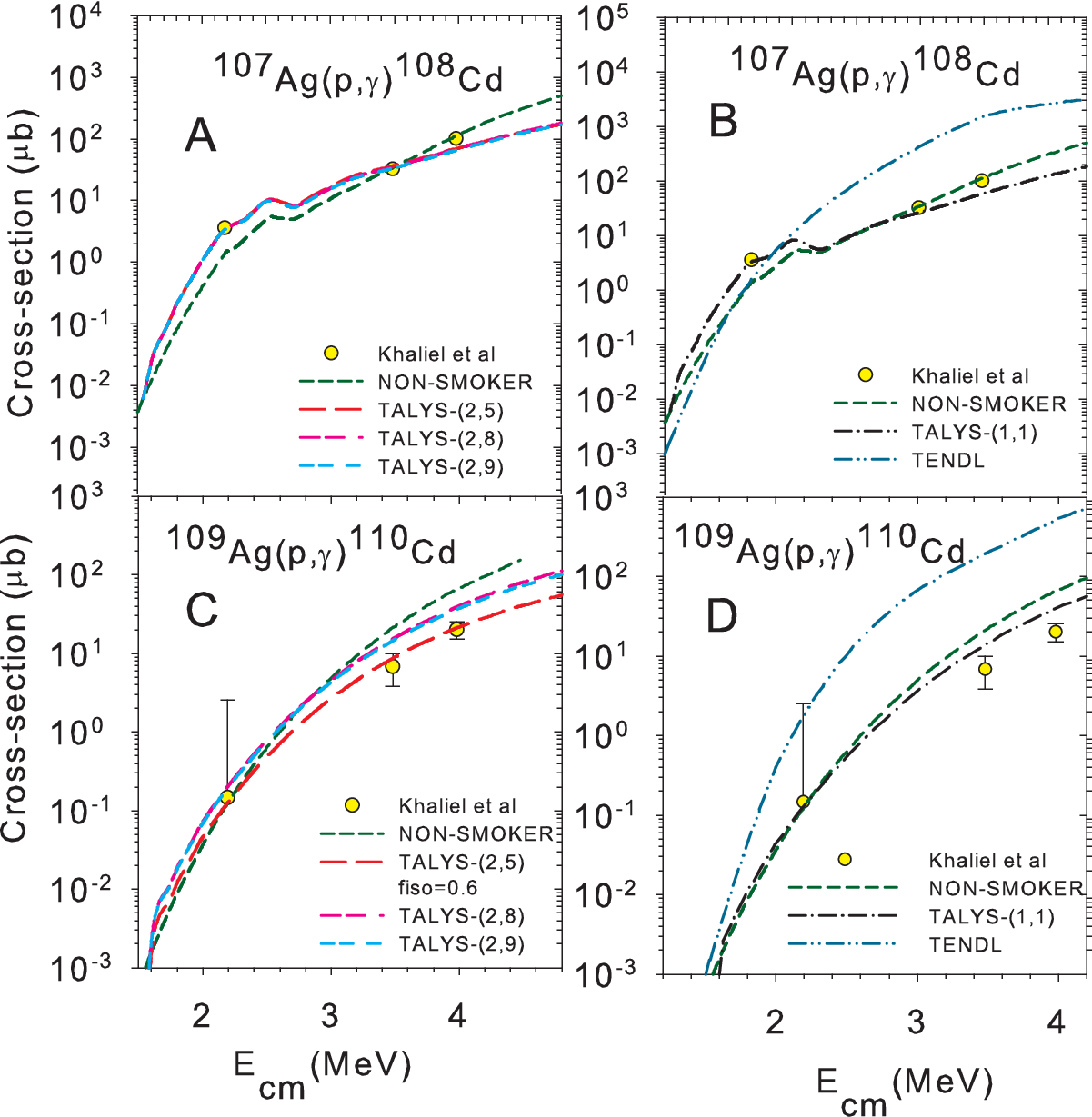}
\caption{\label{fig4} (Online color) Panel A gives the experimental cross-section (Khaliel et al) for the nucleus $^{108}$Cd using yellow filled circles. TALYS-(2,5), TALYS-(2,8) and TALYS-(2,9) (mac-mic) predictions for $^{108}$Cd are shown using the similar plotting conventions as given in earlier figures. Panel B shows TALYS-(1,1) and TENDL plots.
Panels C and D are the corresponding plots for $^{110}$Cd. However, in all the panels corresponding NON-SMOKER plots are also shown.}
\end{center}
\end{figure}

\begin{figure}[htbp]
\begin{center}
\includegraphics[height=8.0 cm, width=8.0cm]{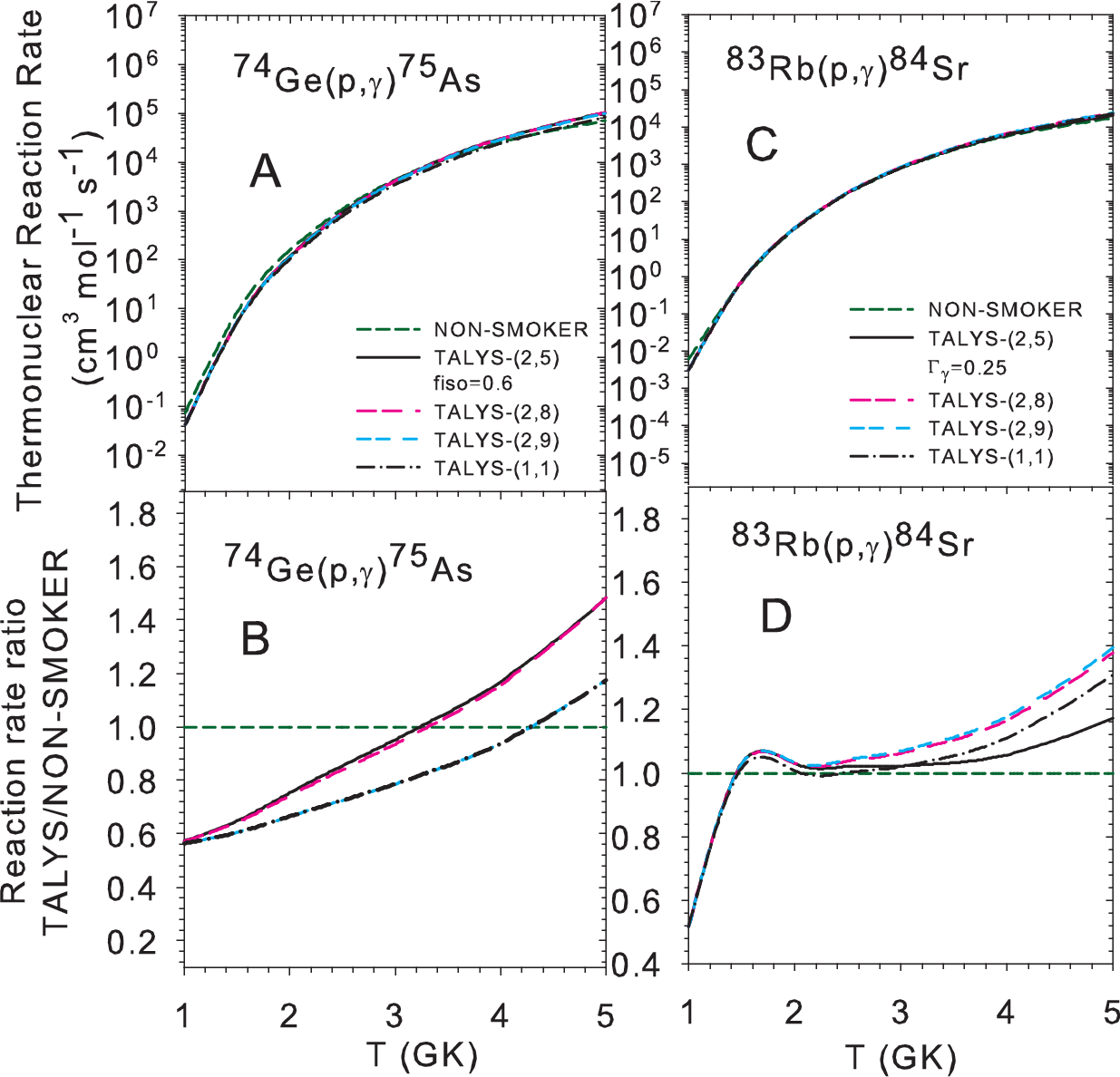}
\caption{\label{fig5} (Online color) Panels A and C show the thermonuclear reaction rate with varying temperatures plotted for NON-SMOKER (green short dashed curve), TALYS-(2,5) with fiso=0.6 (black continuous curve), TALYS-(2,8) (pink medium-medium dashed curve), TALYS-(2,9) (cyan short-short dashed curve), TALYS-(1,1) (black dotted-dashed curve), for the reactions $^{74}$Ge(p,$\gamma$)$^{75}$As and $^{83}$Rb(p,$\gamma$)$^{84}$Sr, respectively. Panels B and D give the corresponding TALYS plots normalized by NON-SMOKER within T=1-5 GK. The green short dashed line denotes NON-SMOKER normalized by itself and hence showing a straight line moving through 1.0.}
\end{center}
\end{figure}

\begin{figure}[htbp]
\begin{center}
\includegraphics[height=8.0 cm, width=8.0cm]{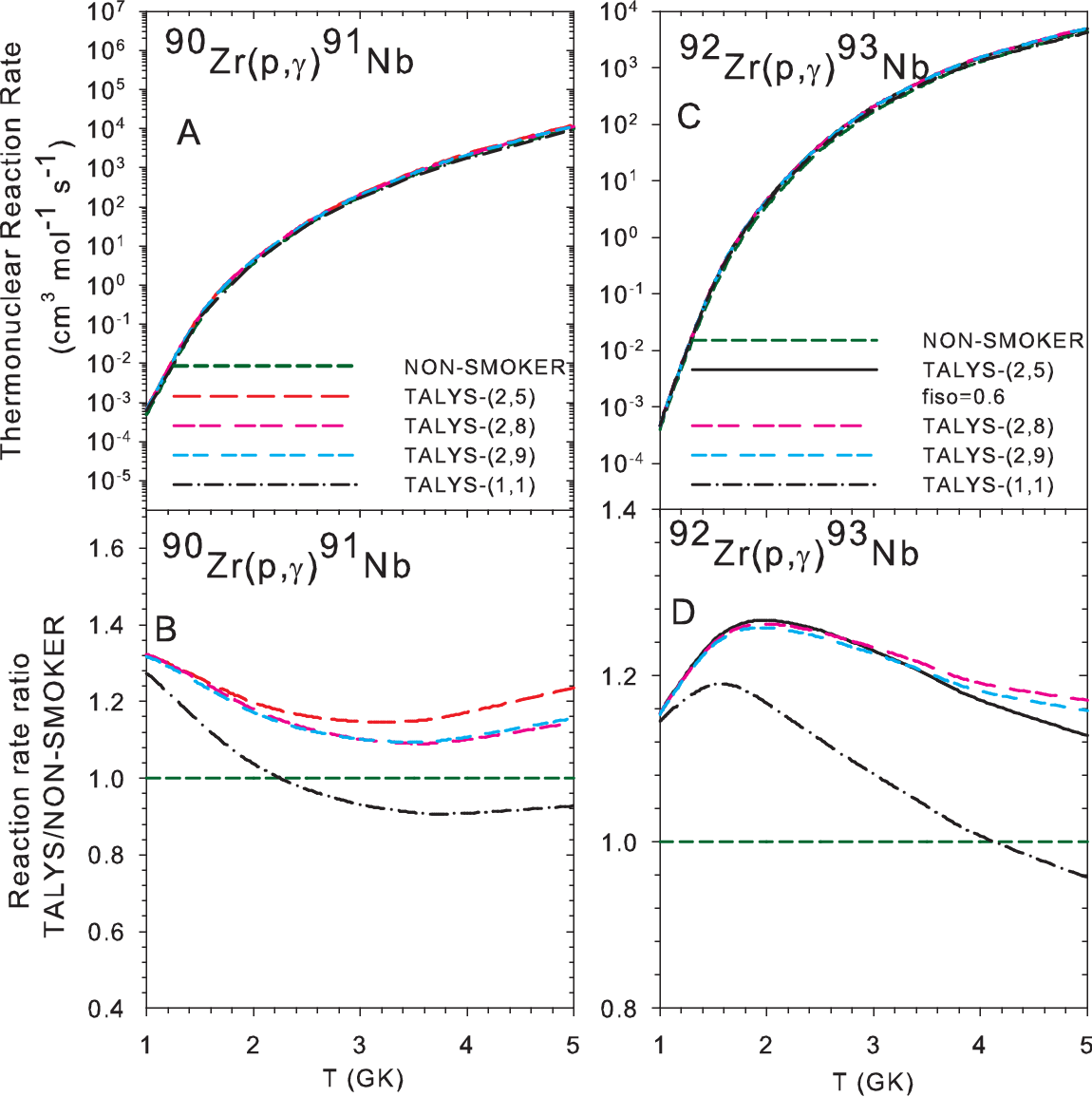}
\caption{\label{fig6} (Online color) Panels A and C show the thermonuclear reaction rate with varying temperatures plotted for NON-SMOKER and different TALYS inputs (as per the conventions followed by earlier Fig. \ref{fig5}) for the reactions $^{90}$Zr(p,$\gamma$)$^{91}$Nb and $^{92}$Zr(p,$\gamma$)$^{93}$Nb, respectively. Panels B and D give the corresponding TALYS plots normalized by NON-SMOKER for $^{75}$As and $^{84}$Sr, respectively, within T=1-5 GK. The green short dashed line denotes NON-SMOKER normalized by itself and hence showing a straight line moving through 1.0.}
\end{center}
\end{figure}

\begin{figure}[htbp]
\begin{center}
\includegraphics[height=8.0 cm, width=8.0cm]{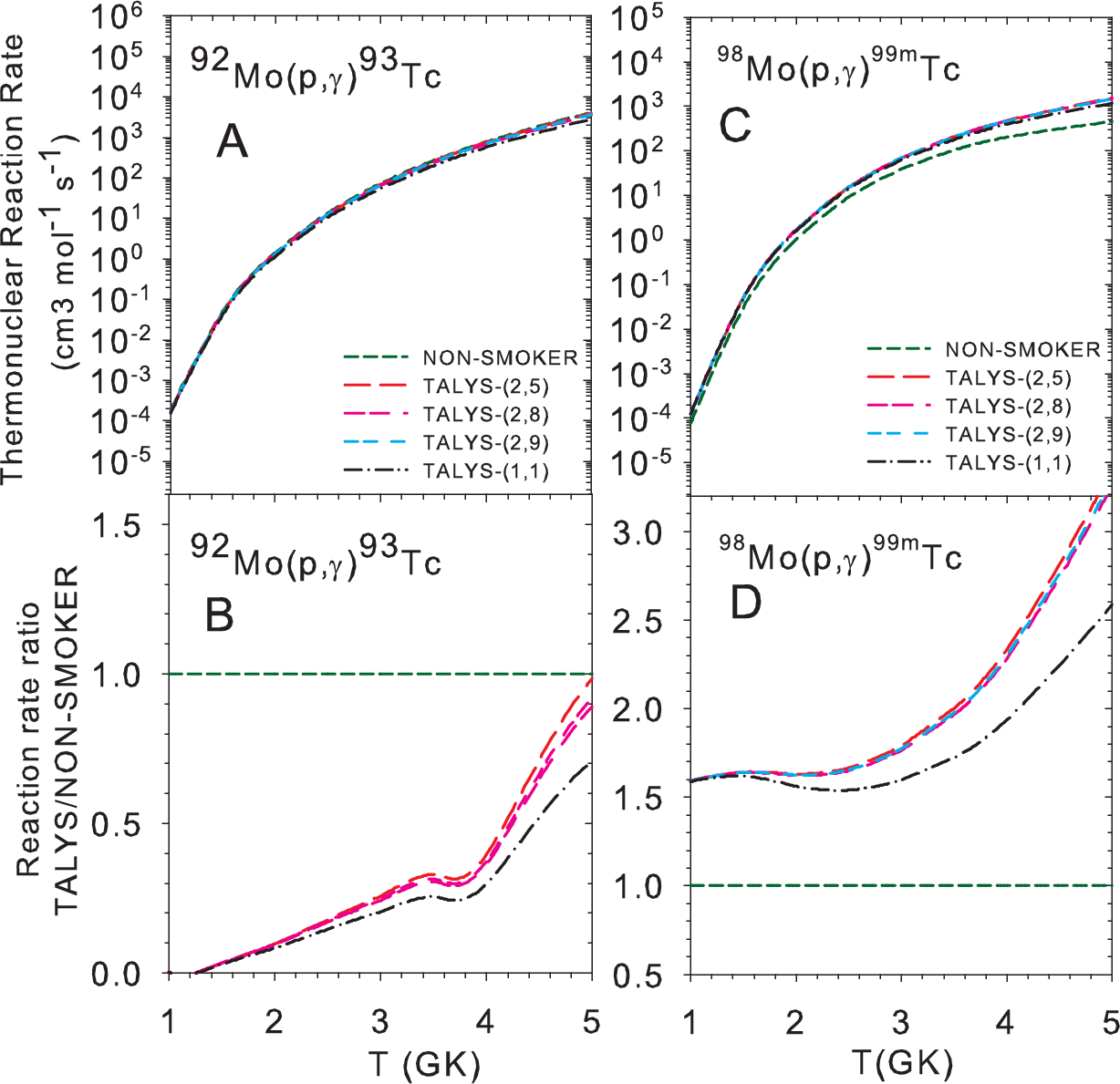}
\caption{\label{fig7} (Online color) Panels A and C show the thermonuclear reaction rate with varying temperature plotted for NON-SMOKER and different TALYS inputs, for the reactions $^{92}$Mo(p,$\gamma$)$^{93}$Tc and $^{98}$Mo(p,$\gamma$)$^{99m}$Tc, respectively. Panels B and D give the corresponding TALYS plots normalized by NON-SMOKER for $^{93}$Tc and $^{99m}$Tc, respectively, within T=1-5 GK. The green short dashed line denotes NON-SMOKER normalized by itself and hence showing a straight line moving through 1.0.}
\end{center}
\end{figure}

\begin{figure}[htbp]
\begin{center}
\includegraphics[height=8.0 cm, width=8.0cm]{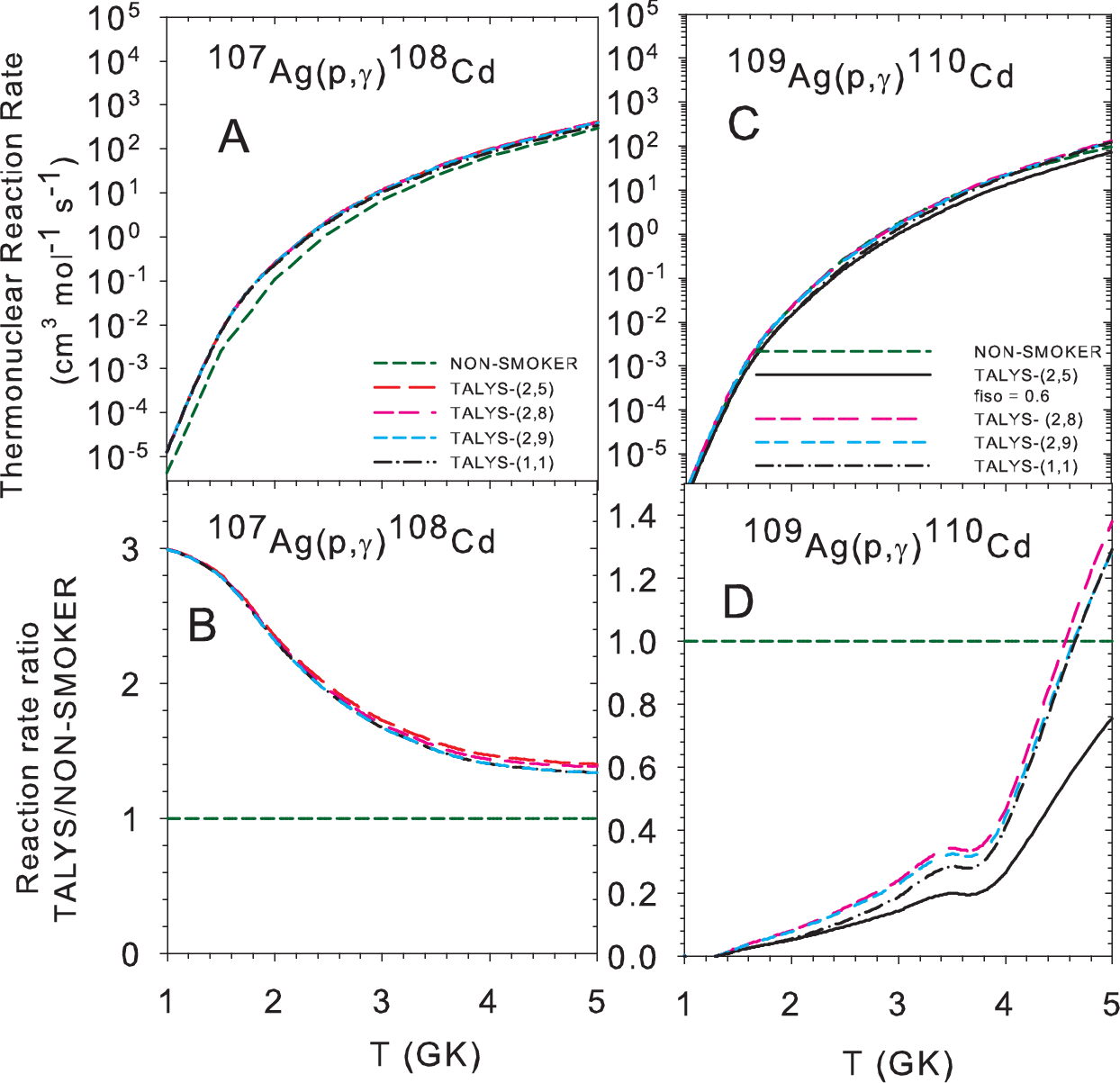}
\caption{\label{fig8} (Online color) Panels A and C show the thermonuclear reaction rate with varying temperature plotted for NON-SMOKER (green short dashed curve), and different TALYS inputs for the reactions $^{107}$Ag(p,$\gamma$)$^{108}$Cd and $^{109}$Ag(p,$\gamma$)$^{110}$Cd, respectively.  Panels B and D give the corresponding TALYS plots normalized by NON-SMOKER for $^{108}$Cd and $^{110}$Cd, respectively, within T=1-5 GK. The green short dashed line denotes NON-SMOKER normalized by itself and hence showing a straight line moving through 1.0.}
\end{center}
\end{figure}

Similarly, for all other nuclides, TALYS calculation was performed and the corresponding $\chi^2$ values were estimated. Panels C and D of Fig. \ref{fig1} demonstrate the TALYS cross-section plots of $^{84}$Sr with mac-mic and mac-mac inputs, while Figs.\ref{fig2}-\ref{fig4} show the corresponding plots for other nuclides, following the same conventions as used in $^{75}$As. 
Panel C of Fig. \ref{fig1} shows cross-section plots for the nucleus $^{84}$Sr with TALYS mac-mic inputs. No theoretical predictions could be found in agreement with the experimental data. Earlier, Williams et al \cite{Will} showed that, after the proton decay width was reduced, SMARAGD code agreed well with the experimental data points. We also observed that reducing $\gamma$-decay width by (1/4)th of its default value, TALYS-(2,5) could also predict the trend of the experimental data. Red long dashed curve is the TALYS-(2,5) prediction, while black continuous line is the same input with reduced $\gamma$ decay width ($\Gamma_{\gamma}=1/4$).    
In panel A of Fig. \ref{fig2}, it can be seen that the TALYS (with mac-mic inputs) plots give similar results predicting the trend of the data of Spyrou et al (yellow circles), while panel B shows TALYS-(1,1) (mac-mac input) output following the trend of the data of Laird et al for the nucleus $^{91}$Nb. For the nucleus $^{93}$Nb, TALYS-(2,5) with fiso=0.6 agrees well with the experimental data, performing better than any other inputs, especially in the higher energy region. The calculations for the nucleus $^{93}$Nb are given in panels C and D.
As shown in the top panels A and B of Fig. \ref{fig3}, the ground state calculations of TALYS-(2,5), (2,8), (2,9) and (1,1) could successfully predict the experimental ground state data in the nucleus $^{93}$Tc. For panels C and D, TALYS-(2,5) agrees well with the experimental data in $^{99m}$Tc compared to other predictions.
Similarly, Fig. \ref{fig4} shows that all TALYS mac-mic calculations could reproduce the experimental data well in $^{108}$Cd, except the lowest energy point, while TALYS-(2,5) successfully reproduces the complete datasets in $^{110}$Cd.
However, corroborated by the above figures, it can be inferred that NON-SMOKER could not represent experimental data accurately in all nuclides, except for $^{108}$Cd.
We have also rechecked the calculation changing the optical model between local and global. The JLM model was also checked on the input. However, we did not obtain any significant deviations.
The corresponding thermonuclear reaction rate plots are also shown in Figs. \ref{fig5}-\ref{fig8} (panels A and C of each figure) for all the nuclides within the temperature T=1-5 GK, which is the region of our interest. The divided plots of TALYS and NON-SMOKER are shown in the panels B and D of the same figures to understand the difference between the predictions. The significant difference between TALYS and NON-SMOKER can be seen from the figures.

 \begin{table}[htbp]
  \begin{center}
    \caption{The TALYS and NON-SMOKER (NS) inputs and corresponding reduced $\chi^2$ and maximum deviation for different nuclei are shown.The minimum value of reduced $\chi^2$ and maximum deviation is given in bold.}
    
       \begin{tabular}{l|c|c|r} 
       
    \hline
       Nucleus & input & reduced $\chi^2$ & max deviation\\
   \hline
$^{75}$As   & \textbf{(2,5)f=0.6} & \textbf{1.4x10$^{-4}$} & \textbf{0.015} \\
                & (1,1)	     & 2.5x10$^{-4}$ & 0.031\\
                & (2,8)	     & 4.2x10$^{-4}$ & 0.107\\
                & (2,9)      & 3.5.x10$^{-4}$ & 0.092\\
                &  NS	     & 0.8420         & 0.612\\
\hline
$^{84}$Sr       & \textbf{(2,5)$\Gamma_{\gamma}$=1/4} & \textbf{26.9} & \textbf{86.9} \\
                & (1,1)	     & 1202.9 & 263.9\\
                & (2,8)	     & 1399.5 & 2699.9\\
                & (2,9)      & 1404.2 & 2750.4\\
                &  NS	     & 311.4  & 428.7\\
\hline

 $^{91}$Nb & \textbf{(2,5)} & 3.16$\times$ 10$^{-5}$  & 3.6$\times$ 10$^{-2}$\\
           & (1,1)	     & 1.5$\times$10$^{-4}$ & 8.0$\times$10$^{-3}$\\
           & (2,8)	     & 2.1$\times$10$^{-5}$ & 3.6$\times$10$^{-2}$\\
           & (2,9)	     & 1.2$\times$10$^{-5}$ & 3.2$\times$10$^{-2}$\\
           &  NS	     & 7.8                  & 6.0$\times$10$^{-1}$\\
\hline
 $^{93}$Nb &	\textbf{(2,5)f=0.6} & \textbf{1.1$\times$10$^{-4}$} & \textbf{3.6$\times$10$^{-2}$}\\
           & (1,1)	     & \textbf{8.2$\times$10$^{-3}$} &\textbf{5.6$\times$10$^{-2}$}\\
           & (2,8)      & 0.2 & 0.1\\
           & (2,9)      & 0.3 & 0.2\\
           &  NS        &\textbf{2.1$\times$10$^{-2}$} & \textbf{6.8$\times$10$^{-2}$}\\
\hline
 $^{93}$Tc & (2,5)      & 29.7	& 657.1\\
           & (1,1)      & 22.6   & 520.4\\
           & (2,8)      & 16.4   & 405.3\\
           & (2,9)	     & \textbf{12.2}   & \textbf{385.3}\\
           &  NS	     & 218.1  &  43.8\\
\hline
$^{99}$Tc & (2,5)      & \textbf{105.7}	& \textbf{57.2}\\
           & (1,1)      & 593.2 & 280.4\\
           & (2,8)      & 230.4 & 103.3\\
           & (2,9)	     & 270.2 & 85.3\\
           &  NS	     & 625.1 & 287.8\\

 \hline 
  $^{108}$Cd & (2,5)      & \textbf{7.6}  & \textbf{41.6}\\
             & (1,1)      & 11.6 & 54.5\\
             & (2,8)	     & 7.9  & 48.2\\
             & (2,9)      & 7.7  & 45.4\\  
            &  NS	     & 8.0   & 42.0\\
\hline

 $^{110}$Cd   & \textbf{(2,5)f=0.6} & \textbf{0.2} & \textbf{0.9}\\
              & (1,1)	              & 0.4          & 1.2\\
             & (2,8) 	 & 9.4 & 9.1\\
              & (2,9) 	 & 9.0 & 8.2\\
             &  NS	     & 217.7 & 208.5\\

\hline

 \end{tabular}
 \label{table}
  \end{center}
\end{table}
 
\begin{figure}
\begin{center}
\includegraphics[height=8.0 cm, width=8.0cm]{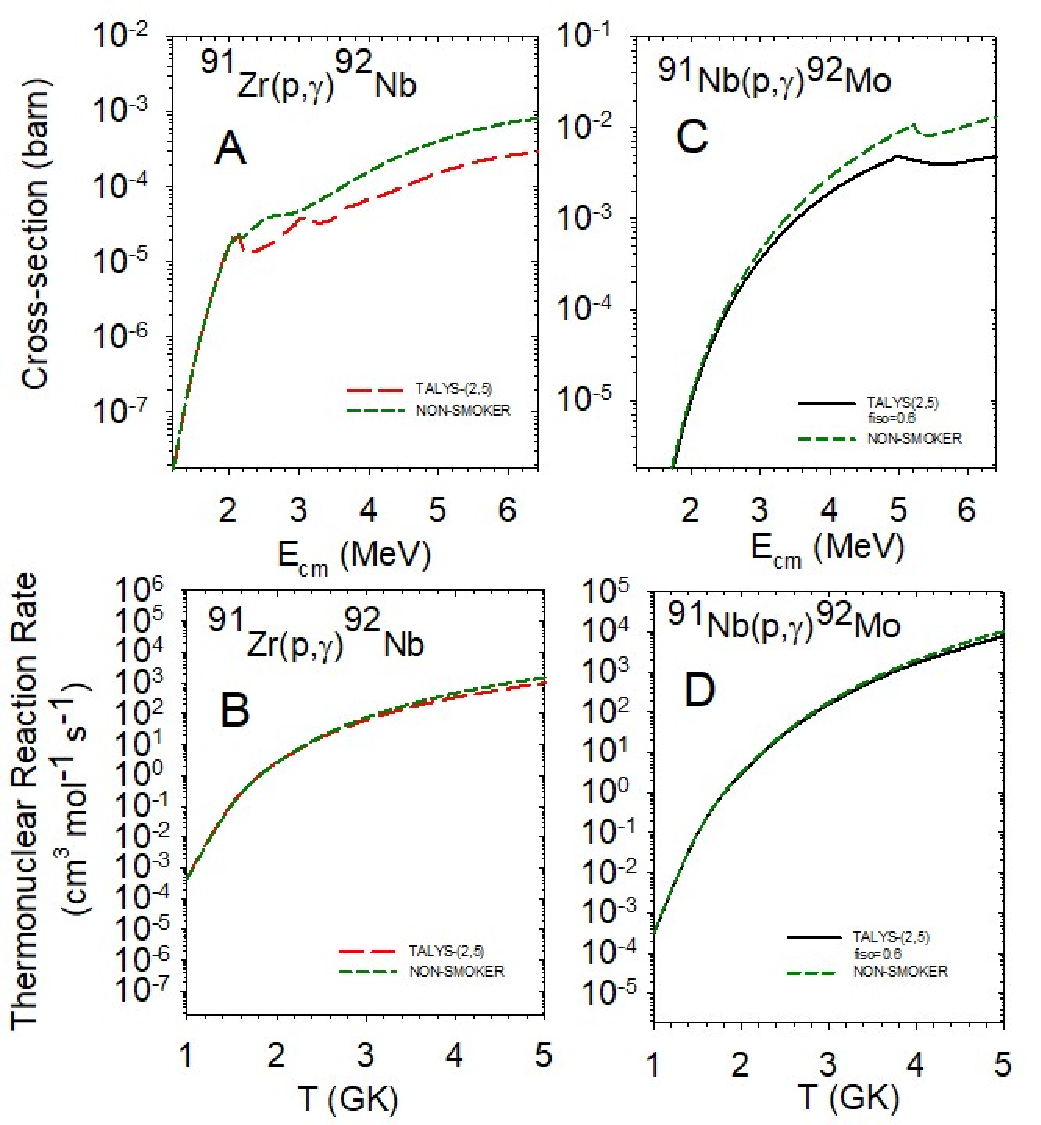}
\caption{\label{fig9} (Online color)  Panel A: TALYS-(2,5) (red long dashed line) and NON-SMOKER calculated (dark green short dashed line) total cross-sections for the nucleus $^{92}$Nb plotted against the centre of mass energy E$_{cm}$ of the corresponding system. Panel B: Thermonuclear reaction rate for the reaction $^{91}$Zr(p,$\gamma$)$^{92}$Nb given by TALYS (2,5) (red long dashed line) and NON-SMOKER (dark green short dashed line), respectively. Panel C: TALYS-(2,5) with isospin correction factor fiso=0.6 (black continuous line) and NON-SMOKER calculated (dark green short dashed line) total cross-sections for the nucleus $^{92}$Mo plotted against the centre of mass energy E$_{cm}$ of the corresponding system.  Panel D: The thermonuclear reaction rate predictions by TALYS-(2,5) with (black continuous line) fiso=0.6 and NON-SMOKER dark green short dashed line) plotted for $^{92}$Mo.}
\end{center}
\end{figure}

The table \ref{table} shows the minimum value of reduced $\chi^2$ and maximum value of deviation for NON-SMOKER and different TALYS inputs. 

This work shows that the code TALYS performes well in this mass region A=75-110, when macroscopic l.d. model and microscopic (or semi-microscopic) $\gamma_{sf}$ (mac-mic) are incorporated in TALYS input. Interestingly, after analyzing $\chi^{2}$ values, the most important point we found that TALYS-(2,5) input combination (l.d parameter and $\gamma_{sf}$ ) performs best for all the nuclei concerned in this work and hence removes the uncertainty in the input models of nuclear reaction code. The better predictive ability of TALYS hybrid (macroscopic l.d.-microscopic $\gamma_{sf}$) inputs against other TALYS inputs implies that the more rigorous microscopic (or semi-microscopic) models perform better than the existing phenemenological models having simplified mathematical formulations of $\gamma$-strength functions. In the recent past, it was seen that reliability of calculations improves if microscopic or semi-microscopic $\gamma$-strength functions are used instead of simplified macroscopic models such as standard Lorentzian (SLO) or Generalized Lorentzian (GLO) \cite{Gor7}. The simplified models usually neglect the lower energy E1 srengths and lower energy M1 mode for deformed nuclei (or in other words, the Scissor mode) and upbending of the M1 deexcitation strength function, utmost important in the lower energy regions (less than 4 MeV or so). Therefore, the microscopic models are more precise. Nevertheless, purely microscopic strength functions sometimes require phenemenological corrections in the splitting of giant dipole resonance strengths. 

The success of TALYS-(2,5) lies in its nature of $\gamma_{sf}$. TALYS(2,5) considers macroscopic high energy Lorentzian formula for E1 strength function, while some of the parameters are microscopically calculated such as the exchange force contribution to sum rule, Migdal constants, etc. Migdal constant denotes the strength of short range nucleon-nucleon interaction. The parameters are validated from experimental data for several nuclei. Therefore it is devoid of errors of purely microscopic as well as macroscopic strength functions. TALYS(2,5) considers detailed microscopic description of photon strength function near the particle decay threshold to include lower energy photon strengths such as E1 strength namely pygmy dipole resonance (PR) and other vibrational modes. One of the characteristic features lies in the inclusion of the strength function dependence on the cube of E$_{\gamma}$ in the nonzero limit of E$_{\gamma}$. Moreover, direct capture processes are included carefully along with the compound nucleus capture. All of these significantly improves the calculation for lower energies. In addition, this also improves the calculation for neutron rich nuclei or other exotic nuclei by describing the PR strengths. The model is successfully tested on 3100 neutron rich nuclei\cite{Gor4}.

We have also included SMLO (option 9) model in this calculation as this is again a semi-microscopic model. This model has been formulated having microscopic mass and deformation dependences of the D1M + QRPA strength and macroscopic Lorentzian type strength function, with a semi-microscopic nature. The performance of SMLO is also found better than macroscopic models \cite{Gor7}. However, inclusion of microscopic l.d model in place of macroscopic l.d model could not improve the performance possibly due to non-adjustment of the parameters with experimental data.
For all the cross-section plots, data from TENDL 23 (TALYS based evaluated nuclear data library version 23) \cite{TEND} are also compared. Except for the nucleus $^{91}$Nb, experimental data have been found not to agree with TENDL. TENDL uses the TALYS code generally with the level density of BSFG and $\gamma_{sf}$ of Kopecky-Uhl (i.e TALYS-(2,2)). 
Regarding the reaction rate, within the region T=1-5 GK, TALYS mac-mic calculations show a large disagreement with the NON-SMOKER results.

\subsection{Application of TALYS-(2,5) to $^{92}$Nb and $^{92}$Mo}

After the selection of correct input combination (TALYS(2,5)), TALYS has finally been used in this work to predict the cross-sections of $^{92}$Nb for the reaction $^{91}$Zr(p, $\gamma$)$^{92}$Nb and $^{92}$Mo for the reaction $^{91}$Nb(p, $\gamma$)$^{92}$Mo, respectively. Both the plots are shown in panels A and B of Fig.\ref{fig9}. 

Earlier, in three ($^{75}$As, $^{93}$Nb and $^{110}$Cd) of the nuclei concerned, we applied isospin correction in terms of the parameter 'fiso'. The value is kept at 0.6 for all these three cases to enhance the neutron channel and to suppress the $\gamma$-ray channel \cite{Lars}. The suppression of $\gamma$-rays just after the threshold energy of the (p,n) channel is evident in the corresponding cross-section plots in comparison to other mac-mic inputs. The energy threshold for the opening of (p,n) reaction channel in the target nucleus $^{74}$Ge is 3.39 MeV. As a result, as we reach E$_{cm}$=3.39 MeV, the isospin- corrected prediction of TALYS-(2,5) shows reduction in the $\gamma$-ray cross-section compared to other TALYS outputs without isospin correction. Similarly, for the target nuclei $^{92}$Zr and $^{109}$Ag, the threshold energy for (p,n) reaction is 2.82 MeV and 2.22 MeV, respectively. Therefore, in these cases, the reduction of $\gamma$-ray cross-section is imminent, which is performed by utilizing the fiso input. So, this work emphasizes the importance of determining the (p,n) reaction threshold energy for the correct analysis of (p,$\gamma$) reactions. For the cases where (p,n) channel energy threshold is within our range of interest, ’fiso’ correction improves the agreement of TALYS with the experiment remarkably.  On the other hand, the reduction of normalization factor in gamma width ($\Gamma_{\gamma}$) could also reduce $\gamma$-ray channel and increase particle decay channels. This is seen in the nucleus $^{84}$Sr.

Therefore, we did not use isospin correction factor in $^{92}$Nb as the (p,n) threshold energy is very low (2.06 MeV). However, isospin correction was performed for $^{92}$Mo as the threshold energy for (p,n) reaction is 5.27 MeV. The corresponding NON-SMOKER predictions are also shown in the same figures. To our knowledge, no previous experimental data exist for both nuclei. The thermonuclear reaction rate of $^{92}$Nb and $^{92}$Mo are also shown in the panels C and D of Fig.\ref{fig9}, respectively. 
Notwithstanding of all the calculations, we could not eliminate the uncertainty arising due to the choice of the parameter 'fiso' or gamma channel width parameter 'gamgamadjust'.  In future, extensive study involving a large number of experimental data is required to remove this uncertainty. 

Here we also did not check the effect of the giant dipole resonance parameters in $\gamma_{sf}$. There are different nuclei which have experimentally measured values of giant dipole resonance (GDR) parameters. It will be interesting to observe how the present TALYS input choice, as mentioned in our work, could perform in the case of nuclei with available experimental GDR parameters in comparison to those without experimental GDR parameters 

\section{Conclusion}
In summary, the matching of TALYS with l.d parameter model of BSFG, $\gamma_{sf}$ of microscopic Goriely table and local OMP (TALYS-(2,5) input set) was found to be best matched with the available experimental data of (p,$\gamma$) capture cross-section for the nuclei with masses ranging between A=75-110. Using this formalism, the agreement between the experimental data and theoretical predictions could be achieved in a uniform way reducing the uncertainty owing to variable input model options. For the nuclei we also add isospin correction that improved the fit in the cases where (p,n) reaction channel opens up. The thermonuclear reaction rates were also extracted from TALYS for the concerned nuclei. The rates, however, do not agree with that of NON-SMOKER theoretical calculations. The same input TALYS-(2,5) is used to find the cross-section of  $^{92}$Nb and $^{92}$Mo, where experimental data are absent. The cross-section of both the nuclei is highly important for the determination of early solar system processes. The reaction rate of both the nuclei is also determined in this work with TALYS-(2,5) input combination. Therefore, in this way, the uncertainties involved in the inputs of nuclear statistical model codes could be removed.

\section{Acknowledgement}
One of the Authors (B.D) acknowledge the research grant of Science and Engineering Research Board (SERB project no: SERB/SUR/2022/000467), Govt. of India.


\begin{thebibliography}{99}
\bibitem{Burb} E. M. Burbidge, G. R. Burbidge, W. A. Fowler, and F. Hoyle. Rev. Mod. Phys. 29(4), 547-650 (1957).
\bibitem{Lang} K. Langanke. Nucl.Phys. A690(1-3), 29C-40C (2001).
\bibitem{Boyd} R. N. Boyd. J.Phys. G-Nucl.Part. Phys. 24(8), 1617-1623 (1998).
\bibitem{Gyur}  G. Gyurky, T.M. Vakulenko. Z. S. Fulop, Z. Halasz, G.G.Kiss, E. Somorjai, and T. Szucs. Nucl. Phys. A922, 112 (2024).
\bibitem{Woos}  S. E. Woosley and W. M. Howard. Astrophys.J.Supp.Ser. 36(2), 285-304 (1978).
\bibitem{Woo1}   S. E. Woosley and W. M. Howard. Astrophys.J. 354(1), L21-L24 (1990).
\bibitem{Raye} M. Rayet, N. Prantzos, and M. Arnould. Astron. Astrophys. 227, 271 (1990).
\bibitem{Raus}  T. Rauscher, N. Dauphas, I. Dillmann, C. Frohlich, Z. Fulop, G. Gyurky. Rep. Prog. Phys. 76, 066201 (2013).
\bibitem{Haus}  W. Hauser and H. Feshbach, Phys. Rev. 87, 366 (1952).
\bibitem{Koni}  A. J. Koning, S. Hilaire, and M. C. Duijvestijn. AIP Conf. Proc. 769, 1154 (2005).
\bibitem{Herm} M. Herman, R. Capote,B.V. Carlson, P. Obložinsky, M. Sin, 
A. Trkov,H. Wienke, V. Zerkin. Nucl. Data Sheets. 108 (2007) 2655.
\bibitem{Rau1}  T. Rauscher, code SMARAGD, version 0.8.4s, 2012..
\bibitem{Rau2}   T. Rauscher and F.-K. Thielemann, At. Data Nucl. Data Tables 75, 1 (2000).
\bibitem{Spyr}  A. Spyrou, S. J. Quinn, A. Simon, T. Rauscher, A. Battaglia, A. Best, et al. Phys. Rev. C88, 045802 (2013).
\bibitem{Khal}  A. Khaliel, T. J. Mertzimekis, E.-M. Asimakopoulou, A. Kanellakopoulos, V. Lagaki, A. Psaltis, et al. Phys. Rev. C96, 035806 (2017).
\bibitem{Wu1} D. Wu, N.Y. Wang, B. Guo, C.Y. He, Y. Tian, X. Tao, et al. Phys. Lett. B805, 135431 (2020).
\bibitem{Mei1} B. Mei, T. Aumann, S. Bishop, K. Blaum, K. Boretzky, F. Bosch, et al.. phys. Rev. C92, 035803 (2015).
\bibitem{Arno} M. Arnould and S. Goriely, Phys. Rep. 384, 1 (2003).
\bibitem{Scha} H. Schatz,A. Aprahamian, J. Görres, M. Wiescher, T. Rauscher, J.F. Rembges, F. K. Thielemann, B. Pfeiffer,P. Möller, 
K. L. Kratz,H. Herndl,B.A. Brown, H. Rebel. Phys. Rep. 294, 167 (1998).
\bibitem{Froh}  C. Frohlich, G. Martínez-Pinedo,M. Liebendorfer,1, F.K. Thielemann, E. Bravo, W. R. Hix, K. Langanke, and N. T. Zinner. Phys. Rev. Lett. 96, 142502 (2006).
\bibitem{Bork}  J. Bork, H. Schatz, F. Kappeler, and T. Rauscher, Phys. Rev. C58(1), 524 (1998).
\bibitem{Lizu}  T. Lizukaa, Y-J. Lai, W. Akram, Y. Amelind, and M. Schonbachler. Earth.Planet.Sci.Lett. 439, 172 (2016).
\bibitem{Tvet} G. M. Tveten, A. Spyrou, R. Schwengner, F. Naqvi, A. C. Larsen, T. K. Eriksen, F. L. Bello Garrote, L. A. Bernstein, D. L. Bleuel, L. Crespo Campo, M. Guttormsen, F. Giacoppo, A.
 Gorgen, T. W. Hagen, K. Hadynska-Klek, M. Klintefjord, B. S. Meyer, H. T. Nyhus, T. Renstrøm, S. J. Rose, E. Sahin, S. Siem, and T. G. Tornyi. Phys. Rev. C94, 025804 (2016).
\bibitem{Lota} G. Lotay, S. A. Gillespie, M. Williams, T. Rauscher, M. Alcorta, M. Amthor, C. A. Andreoiu, D. Baal, G. C. Ball, S. S. Bhattacharjee, H. Behnamian, V. Bildstein, C. Burbadge, W. N. Catford, D. T. Doherty, N. E. Esker, F. H. Garcia, A. B. Garnsworthy, G. Hackman, S. Hallam, K. A. Hudson, S. Jazrawi, E. Kasanda, A. R. L. Kennington, Y. H. Kim, A. Lennarz, R. S. Lubna, C. R. Natzke, N. Nishimura, B. Olaizola, C. Paxman, A. Psaltis, C. E. Svensson, J. Williams, B. Wallis, D. Yates, D. Walter, B. Davids. Phys. Rev. Lett. 127, 112701 (2021).

\bibitem{Will}   M. Williams et al. Phys. Rev. C.107, 035803 (2023)
\bibitem{Lair}  C. E. Laird, D. Flynn, R. L. Hershberger, and F. Gabbard. Phys. Rev. C35, 1265 (1987).
\bibitem{Chen} H. Cheng, B-H. Sun, L-H Zhu, M. Kusakabe, Y. Zheng, L-C. He, et al.. Astrophys. J. 915, 78 (9pp) (2021)..
\bibitem{Gilb} A. Gilbert and A. G. W. Cameron, Can. J. Phys. 43, 1446 (1965).
\bibitem{Dilg} W. Dilg, W. Schantl,H. Vonach,M. Uhl. Nucl. Phys. A217, 269 (1973).
\bibitem{Koni1}   A. J. Koning, S. Hilaire, and S. Goriely, Nucl. Phys. A810, 13 (2008)
\bibitem{Igna} A. V. Ignatyuk, K. K. Istekov, and G. N. Smirenkin, Sov. J. Nucl.Phys. 29, 450 (1979).
\bibitem{Gor1} S. Goriely, F. Tondeur, J.M. Pearson. Atom. Data Nucl. Data Tables 77, 311 (2001).
\bibitem{Gor2}   S. Goriely, S. Hilaire and A.J. Koning. Phys. Rev. C78, 064307 (2008).
\bibitem{Hila}   S. Hilaire, M. Girod, S. Goriely and A.J. Koning. Phys. Rev. C86, 064317 (2013).
\bibitem{Brin}  D.M. Brink, Nucl. Phys. 4, 215 (1957); P. Axel, Phys. Rev. 126, 671 (1962).
\bibitem{Kope} J. Kopecky and M. Uhl, Phys. Rev. C 41, 1941 (1990).
\bibitem{Gor3}  S. Goriely, S. Hilaire, A.J. Koning, M. Sin and R. Capote. Phys. Rev. C79, 024612 (2009).
\bibitem{Gor4}  S. Goriely, Phys. Lett. B436, 10 (1998).
\bibitem{Mart} M. Martini, S. Péru, S. Hilaire, S. Goriely, and F. Lechaftois. Phys.Rev.C94, 014304 (2016).
\bibitem{Gor5} S. Goriely, S. Hilaire, S. Péru, M. Martini, I. Deloncle, and F. Lechaftois. Phys.Rev.C94, 044306 (2016).
\bibitem{Gor6} S. Goriely, S. Hilaire, S. Péru, and K.Sieja, Phys. Rev. C98, 014327 (2018). 
\bibitem{Gor7} S. Goriely and V. Plujko.Phys. Rev. C99, 014303 (2019).
\bibitem{Baug}  E. Bauge, J.P. Delaroche, M. Girod, Phys. Rev. C 63, 024607 (2001); J.P. Jeukenne, A. Lejeune, and C. Mahaux, Phys. Rep. 25C, 83 (1976).
\bibitem{Lars}  A. C. Larsen et al. Phys. Rev. C85, 014320 (2012).

\bibitem{Quin} S. J. Quinn, A. Spyrou, A. Simon, A. Battaglia, M. Couder, and P. A. DeYoung. Phys. Rev. C88, 011603(R) (2013).
\bibitem{Saur}  A. Sauerwein, J. Endres, L. Netterdon, A. Zilges, V. Foteinou, G. Provatas, et al.. Phys. Rev.C 86, 035802 (2012).

\bibitem{TEND} A.J. Koning,D. Rochman,J.Ch. Sublet,N. Dzysiuk, 
M. Fleming, S. van der Marck. Nucl. Data Sheets 155, 1-55 (2019).
 
            


\end{thebibliography}
\end{document}